\documentstyle[times,epsfig]{mn}

\newbox\grsign \setbox\grsign=\hbox{$>$} \newdimen\grdimen \grdimen=\ht\grsign
\newbox\simlessbox \newbox\simgreatbox \newbox\simpropbox
\setbox\simgreatbox=\hbox{\raise.5ex\hbox{$>$}\llap
     {\lower.5ex\hbox{$\sim$}}}\ht1=\grdimen\dp1=0pt
\setbox\simlessbox=\hbox{\raise.5ex\hbox{$<$}\llap
     {\lower.5ex\hbox{$\sim$}}}\ht2=\grdimen\dp2=0pt
\setbox\simpropbox=\hbox{\raise.5ex\hbox{$\propto$}\llap
     {\lower.5ex\hbox{$\sim$}}}\ht2=\grdimen\dp2=0pt
\def\gtrsim{\mathrel{\copy\simgreatbox}}
\def\lesssim{\mathrel{\copy\simlessbox}}

\newcommand{\bez}{\begin{eqnarray*}}
\newcommand{\eez}{\end{eqnarray*}}
\newcommand{\be}{\begin{equation}}
\newcommand{\ee}{\end{equation}}
\newcommand{\beq}{\begin{eqnarray}}
\newcommand{\eeq}{\end{eqnarray}}
\newcommand{\bc}{\begin{center}}
\newcommand{\ec}{\end{center}}

\def\sigmat{\sigma_{\sc T}}
\def\sigmasb{\sigma_{\sc SB}}
\def\taut{\tau_{\rm T}}

\def\taumin{\tau_{\min}}
\def\taumax{\tau_{\max}}
\def\tauave{\overline{\tau}}
\def\cminvsq{\,{\rm cm}^{-2}}
\def\ms{\,{\rm ms}} 
\def\sec{\,{\rm s}}
\def\secinv{\,{\rm s}^{-1}}
\def\eV{\,{\rm eV}}
\def\erg{\,{\rm erg}}
\def\keV{\,{\rm keV}}
\def\Hz{\,{\rm Hz}}
\def\fflat{{f_{\rm flat}}}
\def\fbr{{f_{\rm br}}}
\def\lhmax{l_{h,\max}}
\def\Lhmax{L_{h,\max}}
\def\ls{l_{s}}
\def\lh{l_{h}}
\def\lso{l_{s,0}}

\def\prl{Phys. Rev. Lett.}
\def\pra{Phys. Rev. A}
\def\apj{ApJ}
 
\def\apjl{ApJ}
\def\aap{A\&A}
\def\mnras{MNRAS}
\def\pasj{PASJ}

\begin{document}

\title[Magnetic flares and time lags in accreting black holes]
{Spectral evolution of magnetic flares and 
time lags in accreting black hole sources}

\author[Juri Poutanen and Andrew C. Fabian]
{\parbox[]{6.8in} {Juri Poutanen$^{1,2\star}$ 
and Andrew C. Fabian $^{2\star}$} \\
$^{1}$Stockholm Observatory, S-133 36 Saltsj\"obaden, Sweden \\
$^{2}$Institute of Astronomy, Madingley Road, Cambridge, CB3 OHA} 

\date{Accepted 1999 April 28. In the original form 1998 November 23}

\maketitle


\begin{abstract}
   We  present  a model  for  the  short  timescale  spectral  variability  of
   accreting black holes.  It describes well the time-averaged spectra as well
   as temporal  characteristics  such as  power-density  spectrum,  time/phase
   lags, and coherence function of Cygnus X-1.  We assume that X/$\gamma$-rays
   are produced in compact magnetic flares at radii $\lesssim  100GM/c^2$ from
   the central  black hole.  The tendency  for  magnetic  loops to inflate and
   detach from the underlying accretion disc causes the spectrum of a flare to
   evolve from soft to hard due to the decrease of the feedback  from the cold
   disc, so causing time delays  between  hard and soft  photons.  We identify
   the observed time lags with the evolution  timescales  of the flares, which
   are of the order of the Keplerian timescale.  We model the overall temporal
   variability using a pulse avalanche model in which each flare has a certain
   probability to trigger a neighbouring  flare, thus  occasionally  producing
   long  avalanches.  The duration of the  avalanches  determines  the Fourier
   frequencies where most of the power emerges.
\end{abstract}



\section{Introduction}

   The X-ray  emission of many  sources  harbouring  a black hole shows strong
   variability, on time-scales of  milliseconds  to hours for objects like Cyg
   X-1 and other  Galactic  black  holes  (GBHs),  and  minutes to days for an
   active  galaxy  like  MCG--6-30-15.  The  time-averaged  spectra  are often
   interpreted  as the result of the  Comptonization  of soft photons in a hot
   static electron cloud (e.g.  Shapiro,  Lightman \& Eardley 1976; Sunyaev \&
   Tr\"umper 1979).  Surprisingly, even modern  X/$\gamma$-ray  data of one of
   the best studied bright sources, Cyg X-1, are relatively  well explained in
   terms of such models  (Gierli\'nski  et al.  1997;  Zdziarski et al.  1998;
   Poutanen 1998), despite the observed rapid  variability.  Detailed  studies
   of the X-ray colours,  however, have shown rapid spectral changes (Nolan et
   al.  1981; Negoro et al.  1994), which imply the presence of rapid  changes
   in the physical conditions of the source.

\footnotetext{$\star$ E-mail: juri@astro.su.se (JP);
acf@ast.cam.ac.uk (ACF)}

   The properties of the observed  variability are somewhat  puzzling.  In the
   case of Cygnus X-1, the power-density spectrum (PDS) shows that most of the
   power emerges at Fourier frequencies  corresponding to time-scales  between
   0.1 -- 5~s  (Belloni \& Hasinger  1990; van der Klis 1995).  Studies of the
   autocorrelation function revealed also the existence of strong correlations
   for lags up to $\sim  5\sec$  (Nolan et al.  1981;  Meekins  et al.  1984).
   These  time-scales  exceed by orders of magnitude the  characteristic  time
   scales of the processes operating close to the black hole.

   Another  intriguing  feature is the  Fourier-frequency-dependent  time lags
   between  hard and soft  photons  (Miyamoto  et al.  1988;  Miyamoto  et al.
   1991;  Cui  et  al.  1997;  Nowak  et  al.  1999).  Since  it is  generally
   believed that  Comptonization is responsible for spectral  formation, these
   lags are sometimes  attributed to the delays expected from  scattering in a
   hot cloud.  Hard  photons are the result of more  scattering  and so emerge
   after, or lag behind,  softer ones (Payne  1980).  In order to explain lags
   of order $\sim 0.1\sec$, a huge hot cloud of size $R\sim 10^{9}-10^{10}$~cm
   is then required.  Comptonization  in a uniform  cloud,  however,  produces
   lags  independent of Fourier  frequency  (Miyamoto et al.  1988).  This led
   Kazanas,  Hua \&  Titarchuk  (1997) to  propose  an  inhomogeneous  density
   distribution  of the hot cloud.  While solving one problem,  however,  this
   model raises  another,  more  serious, one of how  physically  to support a
   cloud of such  huge  size for  which  most of the  energy  is at the  outer
   boundary  while  all  the  dissipation  is at  the  centre.  Besides,  when
   modelling the time lags, it is also often assumed that the Compton cloud is
   static and it is the soft photon  input which  varies,  contrary to what is
   observed (Miyamoto et al.  1991).

   Miyamoto \& Kitamoto (1989) were first to propose the spectral evolution as
   a solution  for these  problems.  However, in their model an ad hoc scaling
   between  time-scales of hard and soft pulses is assumed.  Here we present a
   new model for the spectral  variability of accreting black holes and assume
   that the basic unit of energy release is a flare which defines the shortest
   variability  time-scale.  The flares are  assumed  to be  produced  from an
   active region in a stochastic  manner which we model as a pulse  avalanche.
   Individual  flares  are  assumed  to move away  from the  disc, so that the
   fraction  of the  reprocessed  soft flux that  crosses  the flare  emission
   region  ($\equiv$ ER) decreases with time during the life of a flare.  This
   produces the spectral  variability that causes the time delays between hard
   and soft photons of the order of the flare time-scale.

\section{Physical picture}
\label{sec:phys}

   Magnetic  flares  have long been  considered  as the  source  of the  rapid
   variability of Cyg~X-1 and other similar  accreting black holes (Galeev,
   Rosner \& Vaiana 1979; Haardt,  Maraschi \&  Ghisellini  1994; Stern et al.
   1995; Di Matteo 1998; Beloborodov  1999).  Magnetic fields are amplified by
   the  differential  motion in the disc, rise up into the  corona  where they
   reconnect  and liberate  their energy in a flare.  Exactly how this process
   takes place is not clear, but there are  indications  that they  expand and
   detach from the disc (see Mikic, Barnes \& Schnack 1988; Amari,  Luciani \&
   Tagger  1996;  Romanova et al.  1998).  One expects the  overall  evolution
   time-scale  of the  magnetic  field  configuration  to be of the  order  of
   Keplerian time-scale  (Romanova et al.  1998).  Since most of the energy in
   the accretion  disc is dissipated  within radii  $\lesssim  50R_g$ from the
   central  black hole  (here  $R_g\equiv  2GM/c^2$),  these  time-scales  lie
   between  $\sim 1\ms$ and a few tenths of a second for a $M \sim  10M_\odot$
   black hole.

   The characteristic size of the magnetic loop (giving an upper limit for the
   size of the ER) is of the  order of a  turbulent  cell  size,  equal to the
   half-thickness  of the  accretion  disc  (Galeev  et  al.  1979).  For  the
   innermost part of the accretion disc it is then $R\sim \dot{m}R_g$ (Shakura
   \& Sunyaev 1973), where $\dot{m}\equiv \dot{M}c^2/L_{\rm Edd}\sim 1$ in the
   case of Cyg X-1,  and can be  smaller  if a large  fraction  of  energy  is
   dissipated  above the disc, when the disc  height is reduced  (Svensson  \&
   Zdziarski 1994).

   The  reconnection of the magnetic field lines in one location can force the
   neighbouring field configurations into an unstable state causing subsequent
   reconnections  and,  possibly,  a  long  avalanche.  The  evolution  of  an
   individual avalanche may roughly parallel that of a Solar active region (Lu
   \& Hamilton 1991).

\section{Spectral evolution of a magnetic flare} 
\label{sec:evol}

   We follow here the scenario where magnetic  fields are moving away from the
   underlying  accretion  disc in the course of the flare and estimate how the
   flare  spectrum  varies in time.  We assume  that a heating  rate  $\lh(H)$
   ($l\equiv  L\sigmat/(Rm_ec^3)$ is the compactness parameter, and $L$ is the
   luminosity)  is a  function  of  height,  $H$  (measured  in  units  of the
   characteristic size of the ER, $R$), and is zero both when the ER is at the
   surface of the disc and when it is far away, reaching maximum in between.

   The hot electrons  produced by the dissipation are cooled by Comptonization
   of intrinsic soft photons produced in the disc by viscous dissipation (with
   corresponding  compactness  $\lso$) and photons produced by reprocessing in
   the disc of flare  radiation.  In the beginning of the flare, when $\lh$ is
   smaller than $\lso$, the soft  photons are just the average  background  of
   luminosity  $\lso$,  and the  spectrum  emitted  by the flare is very soft.
   When $\lh$ is large, the hard  X-rays  produced  in the ER heat up the cold
   disc in the  vicinity of the flare, so creating  extra soft  photons  which
   then dominate the cooling.  The emitted spectrum defined by the geometry of
   the system is hard (Stern et al.  1995; Svensson  1996).  The luminosity of
   the soft photons that crosses the ER is  $\ls(H)=\lso+  D(H)  \lh(H)$.  The
   height  (and time)  dependent  feedback  factor,  $D(H)$,  determines  what
   fraction of the hard luminosity  after  reprocessing in the disc returns to
   the ER.  During the whole {\it  feedback  phase} when  $\ls(H)\gg\lso$  the
   emission region moves away from disc thus decreasing $D(H)$.  This causes a
   continuous    increase    of    the    Compton    amplification     factor,
   $A\equiv\lh(H)/\ls(H)$,  and a corresponding hardening of the spectrum.  In
   the very last phase, when  intrinsic  disc photons  dominate  cooling,  the
   spectrum becomes soft again.

\begin{figure*}
\centerline{\epsfig{file=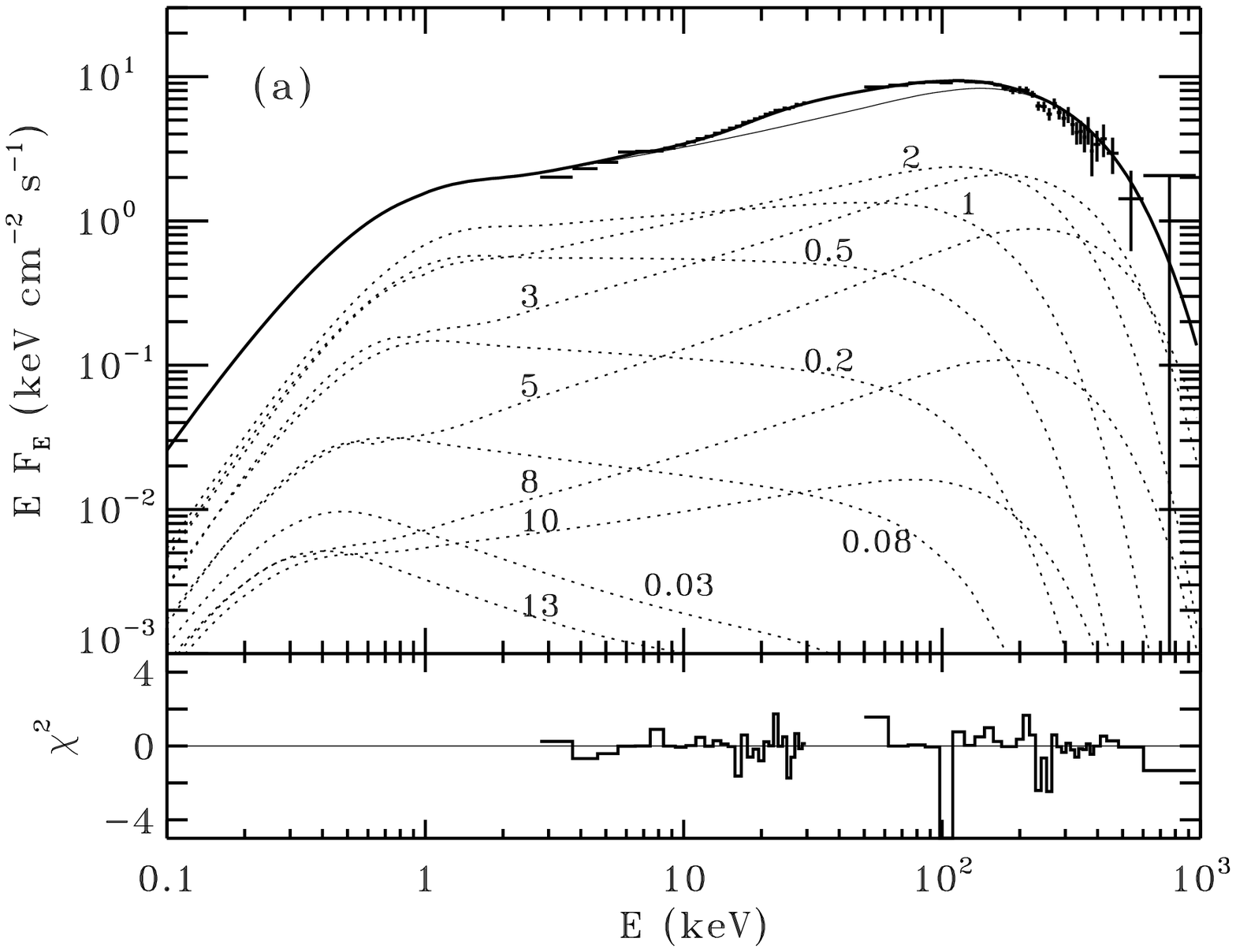,width=9cm,height=6.5cm}
\epsfig{file=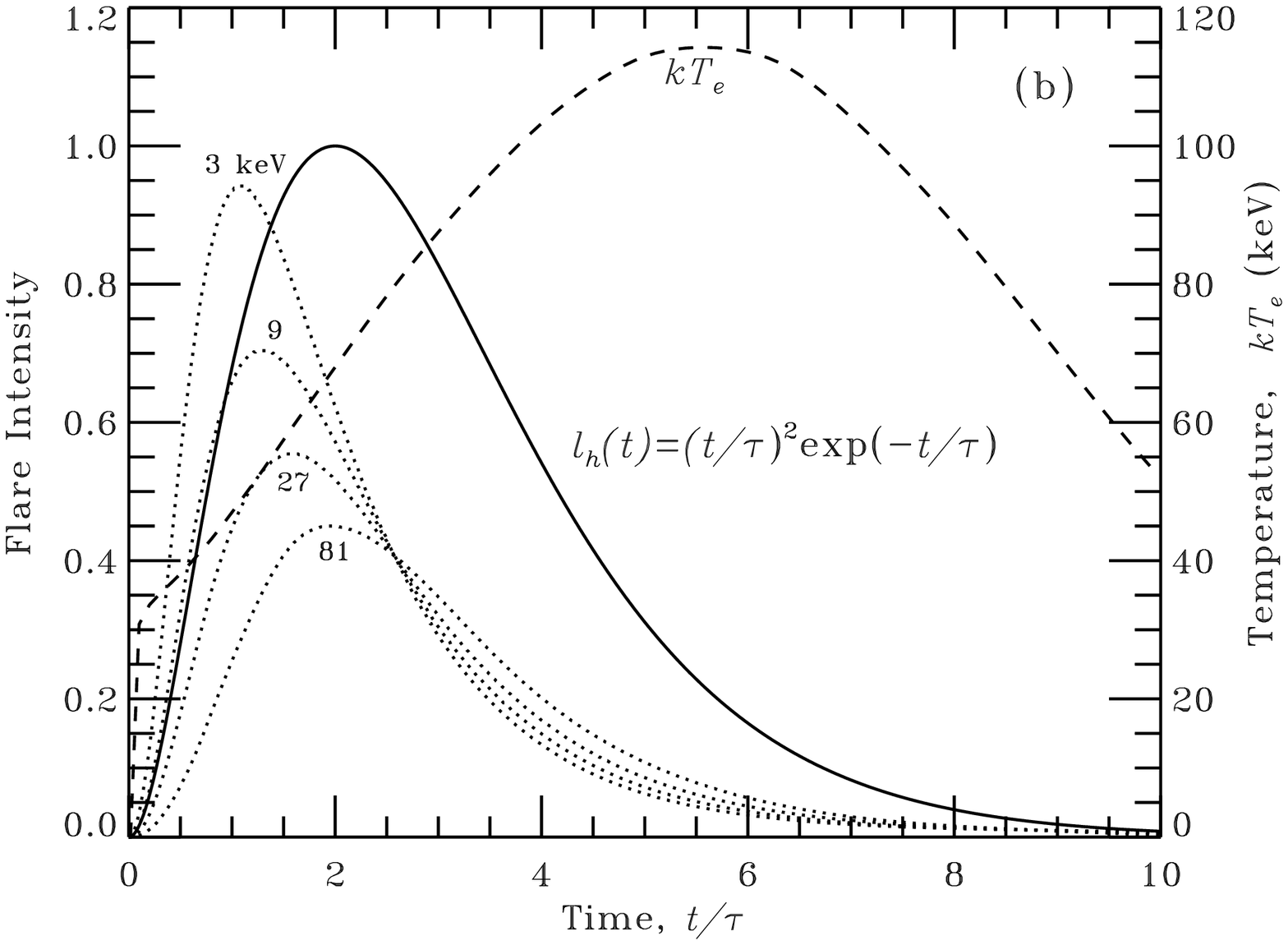,width=8.5cm,height=6.5cm}}
\caption{
   (a) Spectral evolution of a magnetic flare.  Time resolved spectra
   (without Compton reflection) are presented by {\it dotted} curves; marks
   are times in units of $\tau$ from the beginning of the flare.  Time-averaged 
   Comptonized spectrum plotted by a {\it thin solid} curve corresponds to the 
   Compton amplification $A\sim 10$. The Compton reflection is computed by convolving 
   the flare spectrum with the angle-dependent Green functions of
   Magdziarz \& Zdziarski (1995) with inclination fixed at $30^{\rm o}$.
   Time-averaged spectrum of Cygnus X-1
   (simultaneous {\it Ginga} and OSSE data from June 1991, data set 1 in
   Gierli\'nski et al.  1997) is plotted by crosses and the best fit with
   the flare model by a {\it solid} curve ($\chi^2/$dof $=50.0/75$). 
   Interstellar absorption is removed when plotting the model spectrum.  The
   best fit model parameters, $D_0=0.50^{+0.02}_{-0.02}$ and 
   $\taut=1.8^{+0.1}_{-0.1}$. 
   Here for spectral fits we use XSPEC v10 (Arnaud 1996), 
   and the errors are given for a 90 per cent confidence interval 
   ($\Delta \chi^2=2.7$). The amount of Compton reflection 
   $\Omega/(2\pi)=0.30^{+0.03}_{-0.03}$, the EW of Fe line is 100 eV, and 
   absorber column density $N_H=(0.9^{+0.2}_{-0.2})\cdot 10^{22}\cminvsq$.  
   The contribution to the $\chi^2$ is shown at the bottom panel. 
   (b) The flare light curves (in
   arbitrary units) at 3, 9, 27, and 81 keV  are presented by {\it dotted} 
   curves;  the heating rate, $\lh(t)$, by {\it solid} curve.
   Temperature of the ER, $kT_e$ ({\it dashed} curve), peaks at the time
   when Compton amplification $A(t)$ is largest.  Pulses at different
   energies are almost self-similar, i.e.  can be represented by the same
   $(t/\tau_E)^2 \exp(-t/\tau_E)$ law with time-constants  
    at photon energy $E$ approximately following the
   logarithmic dependence, $\tau_E\approx \tau [ 1 + \frac{1}{8} \ln (E/E_p)
   ]$, where $\tau$ is the time-constant of the energy dissipation $\lh(t)$
   and $E_p$ is the photon energy where $EF_E$ peaks.  Thus the delay
   between photons of energies $E$ and $E_0$ is $\Delta t\approx
   \frac{\tau}{4}\ln(E/E_0)$. } \label{fig:evol}
\end{figure*}

   A hardening of the spectrum during the phase of maximal energy  dissipation
   agrees well with the observation of short time-scale  spectral evolution of
   Cyg X-1 (Negoro et al.  1994).  The  luminosity  in hard photons  reaches a
   maximum  later  than the soft  luminosity  does.  In order to show that the
   model  described  above  can  qualitatively  be  applied  to GBHs,  we have
   computed the temporal evolution of the flare spectrum and the time-averaged
   spectrum  using the code of Coppi  (1992) and  compared it with that of Cyg
   X-1.  For an ER of  $R\lesssim  R_g$, we can assume  that the  spectrum  is
   produced  instantaneously  (i.e.  there are no time  delays  due to Compton
   upscattering).  This is a good  approximation  as  long  as  characteristic
   time-scale of the flare $\tau\gg R/c$.  The spectral  evolution then can be
   described as a sequence of steady states.  At any given moment the spectrum
   is defined by three  parameters:  $\ls$, $\lh$, and Thomson   optical depth
   of the ER, $\taut$.  The Compton reflection component is also assumed to be
   produced in the  vicinity  of the flare,  thus  varying  together  with the
   Comptonized radiation.

   The feedback factor can be parametrized as  $D(H)=D_0/(1+3H^2)$,  where the
   uncertainties  in geometry  of the ER,  structure  of the cold disc and the
   angular  distribution  of the flare  radiation  are  incorporated  into the
   factor $D_0$.  The  temperature of the disc that produces soft seed photons
   for Comptonization also depends on the distance of the ER from the disc:
   \be  \label{eq:tbb}  
   T_s(H)  \approx  \left[
   \frac{m_ec^3}{\sigmat   R  \sigmasb   \pi}   
   \left(   l_{s,0}   +\frac{\eta \lh(H)}{1+3H^2}\right) \right]^{1/4} , 
   \ee      
   and equals to  $T_s(H)=0.25  [\lso + \eta  \lh(H)/(1+3H^2)]^{1/4}$  keV for
   $R\sim 3\cdot 10^{6}$ cm ($\sim R_g$, for a $10 M_{\odot}$  black hole). The
   factor $\eta$ combines the  uncertainties  in the angular distribution of 
   the hard radiation and albedo for the reflection  from the disc.

   Assuming  $\lso=0.01$,  the  temperature of the cold disc between flares is
   $T_{s,0}=80  \eV$.  We fix the  compactness  of a flare at the  peak of the
   energy dissipation at $\lhmax=10$  (Haardt et al.  1994; Beloborodov 1999).
   As long as the ratio $\lhmax/\lso$ is large, the spectrum depends weakly on
   our choice of $\lhmax$ (and  $\lso$), but depends  strongly on the feedback
   and the height  where most of the energy  dissipation  occurs.  We  further
   assume, for illustration, that the heating rate varies with time as $\lh(t)
   \propto (t/\tau)^{\xi}  \exp(-t/\tau)$, and the ER moves away from the disc
   with a constant  velocity, so that $H=H_0  t/\tau$.  The exact shape of the
   flare  profile  (value of $\xi$)  does not  affect  any of the  qualitative
   results.  Hard spectra  observed in GBHs require most  dissipation to occur
   at a height  approximately  equal to the size of the ER (Svensson  1996) if
   the  X-ray  emission  is  isotropic  (i.e.  $\eta\sim  1/2$).  Thus we take
   $\xi=2$ (ad hoc), $H_0=1/\xi=1/2$,  $\eta=1/2$, and fit $D_0$.  Altogether,
   we have only two free  parameters,  $D_0$, and the Thomson  optical  depth,
   $\taut$,  that  determine  the  flare  spectrum.  To  be  specific,   $D_0$
   determines the slope of the spectrum, while $\taut$ determines the electron
   temperature  and the spectral  cutoff.  We note that  parameters  $H_0$ and
   $D_0$ are  strongly  correlated,  i.e.  if one  decreases  $H_0$,  the same
   spectrum  can be  produced  decreasing  $D_0$.  In the case of the  thermal
   plasma (assumed here) and $l\sim 10$, the  contribution  from the $e^{\pm}$
   pairs is negligible for the spectrum observed in Cyg X-1.

   With these assumptions, we compute the spectral evolution of the flare (see
   Fig.~1a) and find that the time-averaged spectrum of the flare gives a very
   good fit to the data of Cyg X-1.  If there  were  anisotropy  breaks in the
   time-resolved  spectra  (Stern et al.  1995) due to the  anisotropy  of the
   soft photon injection into the ER (neglected here), the spectral  evolution
   would  smooth  them  out.  In  the   isotropic   case,  the   time-averaged
   Comptonized  spectrum is slightly concave and generally cannot be described
   by a Comptonized spectrum with a fixed optical depth and temperature.

   From the  spectral  evolution  we can  compute  the flare  light  curves in
   various energy bands  (Fig.~1b).  Due to the  soft-to-hard  evolution, soft
   photons come first, while the hard photon  flux (at $\sim$ 100 keV) follows
   the  evolution  of the  heating  rate and peaks  later.  This delay of hard
   photons  relative to soft ones causes time lags to appear in the  composite
   light curves from an ensemble of flares.

\section{The pulse avalanche model} 
\label{sec:aval}

   GBHs as well as many other complex physical  systems often show $1/f$ noise
   in their PDS (Press  1978).  There is no accepted  theory for the origin of
   this   noise.  One   possibility   is  that  a  system   develops   into  a
   self-organized  critical  state (Bak, Tang \& Wiesenfeld  1987;  Mineshige,
   Ouchi \&  Nishimori  1994;  Mineshige  \& Negoro  1999).  In  general,  the
   correlations  between  signals  (flares) of very different  time scales are
   required.  We  model  such  correlations   using  a  version  of  a  simple
   stochastic pulse avalanche model due to Stern \& Svensson (1996) originally
   invented to describe temporal characteristics of gamma ray bursts.

   We assume  that some  flares  occur  spontaneously  according  to a Poisson
   distribution  with a mean rate of  $\lambda $ flares per second  (as in the
   shot noise model, Terrel 1972).  All flares can stimulate the production of
   flares, the number  following a Poisson  distribution of mean number $\mu$.
   Each   spontaneous   flare  ($\equiv$   pulse)  gives  rise  to  a  ``pulse
   avalanche'',  and the whole light curve is  produced  by an overlap of such
   avalanches.  Stimulated pulses are delayed from the stimulating  pulse by a
   time $\Delta t$ which is drawn from a Poisson  distribution of mean $\alpha
   \tau_1$,  where  $\alpha$ is a delay  parameter  and  $\tau_1$  is the time
   constant  of  the  stimulating  pulse.  This  could  be  interpreted  as  a
   ``reservoir''  effect (Begelman \& de Kool 1991).  A long (i.e.  energetic)
   flare would  drain the system of a larger  amount of stored  energy,  which
   results in a longer time to store up enough energy for a subsequent  flare.
   The  presence of such  reservoirs  is also  supported  by the  observations
   (Negoro et al.  1995).  The time  constant of energy  dissipation,  $\tau$,
   for all pulses  (spontaneous  and  stimulated) is assumed to be distributed
   according   to  a  power   law,   $\rho(\tau)\propto   \tau^{-p}$   between
   $\tau_{\min}$  and  $\tau_{\max}$.  We assume that all pulses have the same
   shape and amplitude as shown in Figure~1 and differ only in time  constant,
   $\tau$.  In total  we have  six  parameters:  $\lambda$,  $\mu$,  $\alpha$,
   $\tau_{\min}$,  $\tau_{\max}$  and $p$, but it is quite easy to navigate in
   the parameter  space.  In order to describe the role of the  parameters, we
   discuss how each of them influence the PDS of a simulated light curve.

\section{Results of simulations}
\label{sec:res}

   We  assume  that  all  flares   produce  the   time-resolved   spectra  and
   energy-dependent   pulses  shown  in   Figures~1a,b.  The  variability  and
   interaction  between  individual  flares are described by a pulse avalanche
   model.  Thus we are now able to simulate the light  curves  (Fig.~2a).  The
   model   parameters    ($\lambda=45$;    $\mu=0.8$;    $\alpha=7$;    $p=1$;
   $\taumin=1\ms$;  and  $\taumax=0.15\sec$)  were selected to match a typical
   PDS of Cyg X-1, which can be  described  as a  doubly-broken  power-law:  a
   flat, white noise, part $f^0$ below some frequency $\fflat$; flicker noise,
   $f^{-1}$, between $\fflat$ and $\fbr$; and ``red'' noise, $f^{-\beta}$ with
   $\beta\sim  1.5-2$, above $\fbr$.  Normally,  $\fflat\sim  0.05-0.1\Hz$ and
   $\fbr\sim 2\Hz$ (Fig.~2b; Belloni \& Hasinger 1990; Nowak et al.  1999).

   We identify  $\fbr$ with the  longest  time-scale  of an  individual  flare
   $\taumax\approx  1/(2 \pi \fbr)$.  We note that flares of duration  $\tau$,
   with the dissipation varying as $(t/\tau)^\xi  \exp(-t/\tau)$,  produce PDS
   $\propto [(2\pi f)^2+(1/\tau)^2]^{-(\xi+1)}$ decaying as $f^{-2(\xi+1)}$ at
   high frequencies and giving contributions to the total PDS in a very narrow
   frequency  interval around $f\sim 1/(2 \pi  \tau)>\fbr$,  so that an actual
   value of $\xi$ affects none of the  qualitative  results.  The absence of a
   break at $f\sim 1/(2\pi\taumin)$,  corresponding to the minimum time scale,
   in the  Cyg  X-1  data,  constrains  $\taumin\lesssim  2  \ms$.  The  flare
   time-scales  can be related to the Keplerian  time-scales  at radii between
   the  innermost  stable  orbit and $\sim  30R_g$.  The low  frequency  break
   $\fflat\sim  0.1\Hz$ is related to the  duration  of the  avalanche:  $\sim
   \alpha\tauave\mu/(1-\mu)\sim  1\sec$, where  $\tauave$ is the averaged over
   $\rho(\tau)$  time-scale  of the  flare.  The  power-law  index  $p$ of the
   time-scale distribution is related to the slope, $\beta=3-p$, of the PDS at
   $f>\fbr$ (e.g.,  Lochner, Swank \& Szymkowiak  1991).  The hardening of the
   PDS with photon energy (Nowak et al.  1999) is not  reproduced by our model
   which predicts the same $\beta$ for all energies.

   Parameters  $\alpha$ and $\mu$  determine  the duration of the  avalanches,
   which, in turn,  define the length and slope of the  flicker  noise part of
   the PDS,  where  most of the power  emerges.  The $1/f$  part of the PDS is
   present for a very broad range of parameters  ($4\lesssim\alpha\lesssim 10$
   and  $0.6\lesssim  \mu\lesssim  0.9$).  The larger  $\alpha$ and $\mu$, the
   longer the avalanches, the smaller  $\fflat$, and the larger the fractional
   root-mean-square   (rms)   amplitude   of   the   variability   (i.e.   the
   normalization  of the PDS).  Increase in both  $\alpha$ and $\mu$, thus can
   account for the  observed  anti-correlation  between  $\fflat$  and the rms
   amplitude  (Belloni \& Hasinger  1990).  The average  number of spontaneous
   pulses,  $\lambda$, is a trivial parameter which defines the rms amplitude,
   rms $\propto 1/\sqrt{\lambda}$ ($\sim 30\%$ in the case of Cyg X-1, Belloni
   \& Hasinger 1990; van der Klis 1995).  The rms amplitude decreases slightly
   with photon energy in agreement with the data (Nowak et al.  1999).

\begin{figure*}
\centerline{\epsfig{file=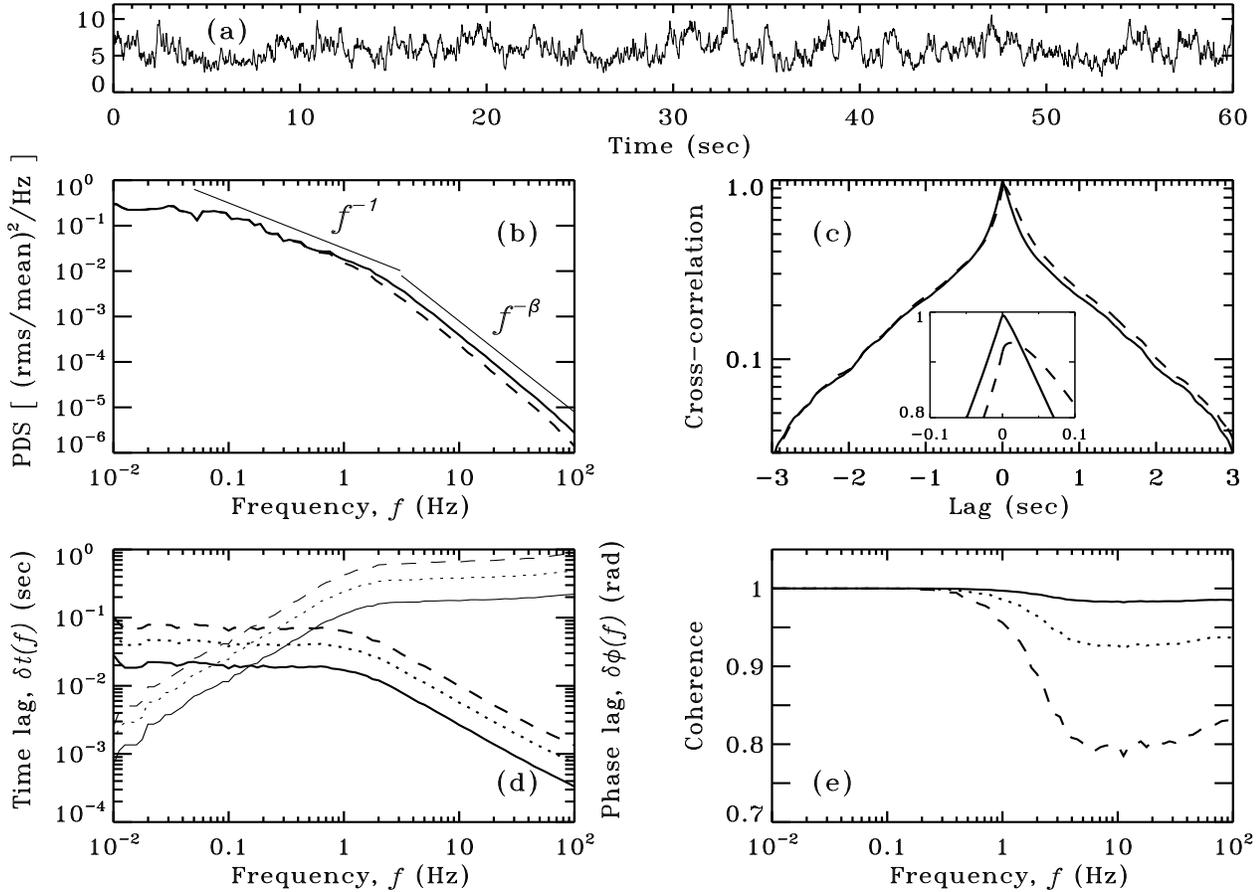,width=17.5cm,height=11.7cm}}
   \caption{(a) An example of a simulated light curve at 3 keV. (b) The
   power-density spectrum (PDS) at 3 keV ({\it solid} curve) and 
   27 keV ({\it dashed}). For a very broad range of
   parameters, the PDS can be represented as a doubly-broken power-law. 
   The slope $\beta$ at high frequencies is related to the slope
   of the time constant distribution, $\beta=3-p$ (in our case $\beta=2$). 
   (c) Normalized cross-correlation functions for 9 keV ({\it solid} curve) 
   and 81 keV ({\it dashed} curve) vs.  3 keV.  The 9 keV CCF peaks at a lag
   $<10^{-3}\sec$ and is almost symmetric; for higher energies, the peak of
   the CCFs shifts more towards positive lags and the CCFs become more
   asymmetric. Insert shows CCF behaviour near the peak. 
   (d) Time lags, $\delta t(f)=\delta \phi(f)/(2\pi f)$ ({\it
   thick} curves), and phase lags, $\delta \phi(f)$ ({\it thin} curves), for
   various energy bands: 9 keV ({\it solid} curve), 27 keV ({\it dotted}),
   and 81 keV ({\it dashed}) vs.  3 keV. Maximum time lags can be
   approximately described by $\delta t_{\max}\approx\frac{\taumax}{8}\ln
   (E/E_0)$, where $E$ and $E_0$ are the corresponding energies and
   $\taumax$ is the maximum flare time-scale. Since time lags are
   proportional to the time-scale of the flare (see
   Fig.~1b), $\delta t(f)\propto \tau \approx 1/(2 \pi
   f)$ (i.e. phase lags are almost constant) at frequencies $f>\fbr$ related
   to individual flares. 
   (e) Coherence functions, with the same definitions as (d). Note the 
   different behaviour at frequencies below and above $\fbr$ and the decrease 
   of the frequency where loss of coherence occurs when $E$ increases. } 
   \label{fig:last}
\end{figure*}

   Besides the PDS, we also compute the normalized cross-correlation functions
   (CCFs),  time/phase lags, and coherence  functions.  The CCFs (Fig.~2c) are
   slightly asymmetric (see, e.g., Nolan et al.  1981) and peak at almost zero
   lags.  They  are   well   described   by  a   stretched   exponential   law
   $\exp(-|t/t_0|^{\nu})$  with the index $\nu\sim 1/2$ and $t_0\sim  0.4\sec$
   within $|t|\lesssim 2\sec$.  A similar stretched  exponential  behaviour is
   found  in the  autocorrelation  function  of  gamma-ray  bursts  (Stern  \&
   Svensson 1996) and is known to describe some  characteristics of turbulence
   (see, e.g., Ching 1991).

   Time lags, $\delta t(f)$  (Fig.~2d),  decay  approximately  as
   $1/f$ above $\fbr$ where  individual  pulses  dominate the PDS:  
   \be 
   \delta t(f)\approx \delta t_{\max} \cdot 
   \left\{  
   \begin{array}{ll}  
   1, & f<\fbr , \\ 
   (f/\fbr)^{-1}, & f>\fbr , 
   \end{array} 
   \right.  
   \ee         
   where $\delta t_{\max} \approx \frac{\taumax}{8} \ln(E/E_0)$ is the maximum
   achievable  time lag  (here  $E$ and  $E_0$  are the  corresponding  photon
   energies).  Since  time  lags are  proportional  to the time  scale  of the
   pulse,  $\tau\approx  1/(2 \pi f)$, which gives  contribution to the PDS at
   frequency  $f$, they are smaller at higher  frequencies.  The  maximum  lag
   $\delta  t_{\max}$ is a factor of two smaller than expected in a shot noise
   model with a fixed time-constant $\taumax$ (see  Fig.~1b)  due
   to  averaging  over   different   time-scales.  A  wide   distribution   of
   time-scales of the individual flares ensures that there are  non-negligible
   time lags and, at the same time, CCFs  peaking  at almost  zero  lags.  The
   logarithmic  dependence  of  time  lags  on  $E$  is  consistent  with  the
   observations  of Cyg X-1 and  other  GBHs  (e.g.,  Cui  1999;  Nowak et al.
   1999).  An amusing fact is that a similar  logarithmic  relation  holds for
   Comptonization  in a static cloud (Payne 1980) where higher energy photons,
   which spend more time in the medium, lag behind  softer ones.  The data for
   Cyg  X-1  (Nowak  et al.  1999)  show  more  complicated  behaviour  of the
   time-lags with Fourier frequency.  Our model reproduces  successfully $1/f$
   dependence  above $\sim 1 \Hz$ and a  ``shoulder''  at $\sim 0.5-1 \Hz$.  A
   break in the time lag at $\sim 0.5 \Hz$ is not reproduced by the model.  In
   \S~6, we speculate on a possible origin of such a break.

   The coherence function (cf.  Bendat \& Piersol 1986; Vaughan \& Nowak 1997)
   is close to unity (Fig.~2e),  since the light curves at different  energies
   are almost perfectly synchronized.  Deviations from unity become visible at
   frequencies    where    individual    pulses    contribute   to   the   PDS
   ($f\gtrsim\fbr\sim  1 \Hz$) and large $E$ which is consistent with the data
   for Cyg X-1 at $f \lesssim 10 \Hz$  (Nowak et al.  1999), but not as strong
   as might be seen for  $f\gtrsim  10\Hz$  (although  we note those  data are
   especially prone to systematic uncertainties that were not fully considered
   by Nowak et al.).  The larger $E$, the smaller the  coherence  function and
   the smaller  Fourier  frequency  where loss of coherence  starts.  The same
   behaviour is expected for any energy dependent shot noise model.

   The parameter $\xi$ of the individual pulse profile affects the time delays
   between  photons of  different  energies.  This, in turn,  affects the time
   lags and the coherence  function in a way that the larger $\xi$, the larger
   the  coherence  and the smaller the time lags.  Changing  $\alpha$,  $\mu$,
   $p$, or  $\lambda$  in the  pulse  avalanche  model  hardly  has {\it  any}
   influence  on the  dependence  of the time lags and  coherence  function on
   frequency.  We note that changing  $\taumax$  affects the break in the PDS,
   $\fbr\sim 1/(2 \pi  \taumax)$,  and  correspondently  changes the frequency
   where loss of coherence  occurs.  The maximum  achievable  time lag is also
   affected, $\delta t_{\max}\propto \taumax$.

   The total averaged  luminosity of the system, $L=\Lhmax \tauave \lambda e^2
   /  [2(1-\mu)]  \erg\secinv$,  determines  the average  peak  luminosity  of
   individual    flares,    $\Lhmax$.   We   find    $\Lhmax\sim   1.6   \cdot
   10^{36}\erg\secinv$    taking    the   Cyg   X-1    luminosity    $L=4\cdot
   10^{37}\erg\secinv$  (Gierli\'nski et al.  1997).  The peak  compactness of
   the  flare  is  then  $\lhmax  \approx  15(R/R_g)^{-1}$.  This  confirms  a
   posteriori our choice of $\lhmax$ in \S~\ref{sec:evol}.  There are $\lambda
   \tauave/  (1-\mu)\sim 7$ flares active at any given time.  The total energy
   dissipated   during   the   flare   $E_{\rm   fl}\sim\Lhmax\overline{\tau}$
   constrains  the  magnetic  field in the  flaring  region to be of the order
   $B\sim \sqrt{6 E_{\rm fl}/R^3} \sim 10^8$~G.

\section{Discussion}
\label{sec:disc}

   The model we present can be developed  in numerous  ways.  We assumed  that
   the spectral  evolution  occurs due to the  inflation of the magnetic  loop
   configuration.  This is, of course, not a unique  possibility.  A number of
   other  scenarios  can produce  similar  behaviour  (see  Poutanen \& Fabian
   1999).  Another assumption that the spectrum is formed  instantaneously  is
   no longer valid for short flares,  $\tau\lesssim$  a few $R/c$ ($\sim 1\ms$
   in our  case,  with a more  stringent  limit  for  higher  observed  photon
   energy),  since  delays  due to  Compton  upscattering  and  production  of
   reprocessed  soft photons then have to be  incorporated.  The delays in the
   production of soft (reprocessed)  photons can make the Comptonized spectrum
   hard in the beginning of the flare and soft later  producing {\it negative}
   lags at $f\gtrsim 50\Hz$.

   In our  model,  the  longer  flares  appear to be very  smooth,  while,  in
   reality, they can consist of a train of correlated,  much shorter,  flares.
   This  effect  can  also  significantly  affect  the PDS  behaviour  at high
   frequencies  (possibly  even  accounting  for the hardening of the PDS with
   photon energy, Nowak et al.  1999).

   Another  simplification  is that we  assumed  that  Compton  reflection  is
   produced in the  vicinity of the flare.  However, it is possible  that some
   fraction of the hard radiation is reflected, e.g., from the outer disc, the
   wind,  or the  companion  star.  This  would  cause  a  suppression  of the
   variability of this  component at short  time-scales  (acting as a low-pass
   filter) and produce additional time lags independent of Fourier  frequency.
   A break in the time lag -- Fourier  frequency  relation  should then appear
   for energies where Compton  reflection is not negligible  (e.g.,  comparing
   $E\gtrsim 7 \keV$ and  $E\lesssim  2\keV$  channels)  and it can be used to
   determine the distance of the reflector  from the X-ray source.  We propose
   that the  break in the time  lags at  $f\sim  0.5\Hz$  observed  in Cyg X-1
   (Nowak et al.  1999) can be a consequence of this effect.

   An important issue is the small amplitude of Compton reflection seen in Cyg
   X-1 and other GBHs in their hard state (Ebisawa et al.  1996;  Gierli\'nski
   et al.  1997; Zdziarski et al.  1998),  $\Omega/(2\pi)\approx  0.3$.  These
   observations have been interpreted as an absence of reflecting  material in
   the direct vicinity of the X-ray source (cf.  Dove et al.  1997;  Poutanen,
   Krolik \& Ryde 1997).  In the present model, we explicitly  assume that the
   cold material exists in the vicinity of the flare and produces soft photons
   by  reprocessing  hard X-ray  radiation.  Possible  solutions  include  the
   following  alternatives:  (a) the cold disc is  disrupted  into a number of
   cold filaments  that produce  enough seed soft photons for  Comptonization,
   but have a too small covering factor to produce visible Compton  reflection
   (Lightman  1974; Krolik 1998;  Zdziarski et al.  1998; Poutanen  1998); (b)
   there is bulk motion  directed away from the disc and more flare  radiation
   is beamed toward the observer than the disc (Beloborodov  1999); or (c) the
   disc is highly ionized (Ross, Fabian \& Young 1999).

   An  interesting  question is why the time lags in the soft state of Cyg X-1
   are  smaller  than  those  in the  hard  state  (Cui et al.  1997).  We can
   suggest at least two reasons.  First, since the soft state spectrum is {\it
   softer}, the amplitude of the spectral  evolution is much smaller.  Second,
   if in the soft state  X/$\gamma$-rays are produced only within $\sim 10R_g$
   from the black  hole, the  dynamical  time-scales  are on average  smaller,
   causing smaller lags.

\section{Summary}
\label{sec:sum}
 
   We have presented a model for the short time-scale spectral  variability of
   accreting  black  holes, and  showed  using  Cygnus  X-1 that it is able to
   describe time-averaged spectra as well as general behaviour of the temporal
   characteristics  such  as  power-density   spectra,  time/phase  lags,  and
   coherence  function.  We based  our {\it  spectral}  model on the fact that
   magnetic  loops have a tendency to inflate  and detach  from the  accretion
   disc.  The corresponding decrease of the feedback from the cold disc causes
   the  soft-to-hard  evolution of the flare spectrum  which is the reason for
   time lags between hard and soft photons.  We assumed that all properties of
   the flares scale with their  time-scale,  $\tau$.  This  produces  the time
   delays  between hard and soft photons in an  individual  flare also scaling
   with  $\tau$.  In their turn, the time lags from an  ensemble  of flares of
   different duration show the $1/f$ dependence on the Fourier frequency.

   The model for the {\it temporal  variability} is based on a pulse avalanche
   model where flares can, with some  probability,  trigger other flares.  The
   duration  of the  avalanches,  which are the  result  of this  interaction,
   determines  the  Fourier  frequencies  where  most  of the  power  emerges.

   We identified the break in the PDS at $\fbr$ with the maximum time-scale of
   the individual flare  $\taumax\approx  1/(2 \pi \fbr)$ and pointed out that
   statistical  properties  are  very  different  at  $f<\fbr$  and  $f>\fbr$.
   Deviation  from  unity  of the  coherence  function  and the  break  in the
   time/phase  lag spectrum  occur also close to $\fbr$.  The PDS,  time/phase
   lags, and  coherence  at  $f>\fbr$  are  related to the  properties  of the
   individual  flares,  while the  behaviour  at $f<\fbr$  is a property  of a
   stochastic  process.  Our model  predicts that  variations of $\fbr$ in the
   PDS should be accompanied by  corresponding  changes in the  time/phase lag
   and coherence spectra.  

   The success of the model in  reproducing  most of the spectral and temporal
   properties   of  Cygnus   X-1   supports   the   point  of  view  in  which
   X/$\gamma$-rays  are produced in localized  hot regions  (magnetic  flares)
   that  interact  with each  other.  We expect  that this  model  may also be
   useful when  analyzing  data from other  accreting  objects  such as active
   galactic nuclei and weakly magnetized neutron stars.

\section*{ACKNOWLEDGMENTS}
 
   We are  grateful  to Boris  Stern for the  original  version  of the  pulse
   avalanche  model and to Paolo  Coppi  for the {\sc  eqpair}  code.  Helpful
   discussions with the participants of the workshop on High Energy  Processes
   in Accreting  Black Holes and useful  suggestions of the anonymous  referee
   are  acknowledged.  JP is grateful to the  Institute of  Astronomy  for the
   hospitality  during  his  visit.  His work  was  supported  by the  Swedish
   Natural  Science  Research  Council and the Anna-Greta and Holger  Crafoord
   Fund.  ACF thanks the Royal Society for support.


\begin{thebibliography}{}

\bibitem[{} {}]{ama}
Amari T., Luciani J.~F., Tagger M., 1996, A\&A, 306, 913 

\bibitem[{} {}]{arn}
Arnaud K.~A., 1996, in Jacoby G.~H., Barnes J., eds, ASP Conf. Ser. Vol. 101, 
Astronomical Data Analysis Software and Systems V. ASP, San Francisco, p. 17

\bibitem[Bak, Tang, \& Wiesenfeld 1987]{bak87}
Bak P., Tang C.,  Wiesenfeld K., 1987,  \prl, 59, 381

\bibitem[{} {}]{beg}
Begelman M.~C., de Kool M., 1991, in Miller H.~R., Wiita P.~J., eds, 
 Variability of Active Galactic Nuclei.  
 Cambridge Univ. Press, Cambridge, p. 198 

\bibitem[{} {}]{bh1}
Belloni T., Hasinger G., 1990, A\&A,  227, L33 

\bibitem[{} {}]{be}
Beloborodov A.~M., 1999, \apjl,  510, L123
 
\bibitem[{} {}]{bp86}
Bendat J.~S., Piersol A.~G., 1986, Random Data: Analysis and Measurement 
Procedures. Wiley, New York 

\bibitem[Ching 1994]{chi94}
Ching E., 1991, \pra, 44, 3622

\bibitem[{} {}]{coppi92}
Coppi P.~S., 1992, \mnras, 258, 657 

\bibitem[{} {}]{cui99}
Cui W., 1999,  in Poutanen J., Svensson R., eds, 
ASP Conf. Ser. Vol. 161, High Energy Processes in Accreting Black Holes. 
ASP, San Francisco, p. 97 (astro-ph/9809408) 

\bibitem[{} {}]{cui97}
Cui W., Zhang S.~N., Focke W., Swank J.~H., 1997, \apj, 484, 383 

\bibitem[{} {}]{dimat1} 
Di Matteo T., 1998, \mnras, 299, L15 

\bibitem[ ]{dove97}
Dove J.~B., Wilms J., Maisack M., Begelman M.~C., 1997, \apj, 487, 759

\bibitem[{} {}]{ebi}
Ebisawa K., Ueda Y., Inoue H., Tanaka Y., White N.~E., 1996, 
ApJ, 467, 419

\bibitem[{} {}]{gal}
Galeev A.~A., Rosner R., Vaiana G.~S., 1979, \apj, 229, 318

\bibitem[{} {}]{gier}
Gierli\'nski M., Zdziarski A.~A., Done C., Johnson W.~N., Ebisawa K.,
  Ueda Y., Haardt F., Phlips B.~F., 1997, \mnras, 288, 958

\bibitem[{} {}]{hmg94}
Haardt F., Maraschi L., Ghisellini G., 1994, \apj, 432, L95

\bibitem[{} {}]{kaz}
Kazanas D., Hua X.-M., Titarchuk L., 1997, \apj, 480, 735 

\bibitem[{} {}]{kro98}
Krolik J.~H., 1998, \apj, 498, L13   

\bibitem[]{lightman74}
Lightman A.~P., 1974, \apj, 194, 429

\bibitem[ ]{lochner}
Lochner J.~C., Swank  J.~H., Szymkowiak A.~E., 1991, \apj, 376, 295 
 
\bibitem[ ]{lu}
Lu E.~T., Hamilton R.~J., 1991, \apj, 380, L89

\bibitem[ ]{mz95}  
Magdziarz P.,  Zdziarski A.~A., 1995, \mnras, 273, 837

\bibitem[ ]{meekins}
Meekins J.~F. et al., 1984, \apj, 278, 297 
       
\bibitem[ ]{mi}
Mikic Z.,  Barnes D.~C., Schnack D.~D., 1988, \apj, 328, 830 

\bibitem[Mineshige \& Negoro (1999)]{mine99}
Mineshige S., Negoro H., 1999, in Poutanen J., Svensson R., eds, 
ASP Conf. Ser. Vol. 161, High Energy Processes in Accreting Black Holes. 
ASP, San Francisco, p. 113
 
\bibitem[Mineshige et al. (1994)]{mine94}
Mineshige S., Ouchi N.~B., Nishimori H., 1994, \pasj, 46, 97  

\bibitem[ ]{miy2}
Miyamoto S., Kitamoto S., 1989, Nature, 342, 773 

\bibitem[ ]{miy1}
Miyamoto S., Kitamoto S., Mitsuda K., Dotani T., 1988, Nature, 336, 450

\bibitem[ ]{miy3}
Miyamoto S., Kimura K., Kitamoto S., Dotani T., Ebisawa K., 1991, 
\apj, 383, 784

\bibitem[ ]{neg}
Negoro H., Miyamoto S., Kitamoto S., 1994, \apj, 423, L127 

\bibitem[ ]{neg2}
Negoro H., Kitamoto S., Takeuchi M., Mineshige S.,   1995, \apj, 452, L49

\bibitem[ ]{nol}
Nolan P.~L. et al., 1981, \apj, 246, 494 

\bibitem[ ]{now}
Nowak M.~A., Vaughan B.~A., Wilms J., Dove J.~B., Begelman M.~C., 1999, 
\apj, 510, 874

\bibitem[ ]{payne} 
Payne D.~G., 1980, \apj, 237, 951 

\bibitem[ ]{pou1}
Poutanen J., 1998, in Abramowicz M.~A., Bj\"ornsson G., Pringle J.~E., eds, 
Theory of Black Hole Accretion Discs. Cambridge 
Univ. Press,  Cambridge, p. 100

\bibitem[ ]{pf99}
Poutanen J., Fabian A.~C., 1999, in Poutanen J., Svensson R., eds, 
ASP Conf. Ser. Vol. 161, High Energy Processes in Accreting Black Holes. 
ASP, San Francisco, p. 135 

\bibitem[]{pkr97}
Poutanen J., Krolik J.~H., Ryde F., 1997,  \mnras, 292, L21

\bibitem[]{press}
Press W.~H., 1978, Comm. Astrophys., 7,  103

\bibitem[ ]{rom}
Romanova M.~M., Ustyugova G.~V., Koldoba A.~V., Chechetkin V.~M.,
Lovelace R.~V.~E., 1998, \apj,  500, 703 

\bibitem[ ]{ross99}
Ross R.~R., Fabian A.~C., Young A.~J., 1999, \mnras, in press (astro-ph/9902325)

\bibitem[ ]{ss73}
Shakura N.~I., Sunyaev R.~A., 1973, \aap, 24, 337

\bibitem[ ]{sle}
Shapiro S.~L., Lightman A.~P., Eardley D.~N., 1976, \apj, 204, 187

\bibitem[ ]{ssv}
Stern B.~E., Svensson R., 1996, \apj, 469, L109 

\bibitem[]{ste95b}
Stern B.~E., Poutanen J., Svensson R., Sikora M.,
   Begelman M.~C., 1995, \apj, 449, L13 

\bibitem[ ]{st79}
Sunyaev R.~A., Tr\"umper J., 1979, Nature, 279, 506 

\bibitem[]{sve96b}
Svensson R., 1996,  A\&AS, 120, 475

\bibitem[]{sz94}
Svensson R., Zdziarski A.~A., 1994, \apj, 436, 599
   
\bibitem[ ]{ter}
Terrel N.~J.~Jr., 1972, \apj, 174, L35 

\bibitem[ ]{klis}
Van der Klis M., 1995, in Lewin W.~H.~G., van Paradijs J., 
  van den Heuvel E.~P.~J., eds,   
  Cambridge Astrophys. Ser. Vol. 26, X-ray Binaries.   
  Cambridge Univ. Press, Cambridge, p. 252   

\bibitem[ ]{vn97} 
Vaughan B.~A., Nowak M.~A., 1997, \apj, 474, L43 

\bibitem[ ]{zdz1}
Zdziarski A.~A., Poutanen J., Miko\l ajewska J.,
Gierli\'nski M., Ebisawa K., Johnson W.~N., 1998, \mnras, 301, 435 

\end{thebibliography}
\end{document}